\documentclass[12pt]{article}

\usepackage{graphicx}

\usepackage{amsmath}
\usepackage{amssymb}
\usepackage[square]{natbib}

\begin{document}

\title{Trouble With The Curve:  Improving MLB Pitch Classification}
\author{Michael A. Pane \and Samuel L. Ventura \and Rebecca C. Steorts \and A.C. Thomas}
\date{\today}

\maketitle

\begin{abstract}

The PITCHf/x  database has allowed the statistical analysis of of Major League Baseball (MLB) to flourish
since its introduction in late 2006. Using PITCHf/x, pitches have been classified by hand, requiring considerable effort, or using neural network clustering and classification, which is often difficult to interpret. To address these issues, we use model-based clustering with a multivariate Gaussian mixture model and an appropriate adjustment factor as an alternative to current methods. Furthermore, we describe a new pitch classification algorithm based on our clustering approach to address the problems of pitch misclassification. We illustrate our methods for various pitchers from the PITCHf/x database that covers a wide variety of pitch types.

\end{abstract}

\section{Introduction}

Perhaps more than any other technological contribution in baseball, the deployment of the PITCHf/x system has proven to be an invaluable resource to teams and to fans of the game in their statistical analyses of baseball, both in its original form and as augmented by estimates of pitch spin parameters (known as ``psuedo-spin'').  End users of the system, particularly brooksbaseball.net and MLB Advanced Media (MLB-AM), have classified these pitches into common pitch types, both by hand and using neural network classification methods. We hypothesize that the current method for classifying pitches in the database can be refined due to evidence from graphical exploration and previous research [1]. \\

We improve the current pitch classification system by comparing the current neural network classification results to these from various statistical clustering methods, such as k-means, hierarchical clustering, and model-based clustering with a multivariate Gaussian mixture model (MBC).  We ultimately propose an alternative basis for pitch clustering and classification. In Section~2 we describe the current methods used and the PITCHf/x database. In Section~3, we introduce the model-based clustering method, which has several advantages for examining pitch types including pitches with high variance, pitch evolution across time, and pitches with similar characteristics. We also address implementation of the MBC method as well as the selection of the clusters and the stability of this clustering method. In Section~4, we propose a novel algorithm for classifying pitches based on their characteristics, before concluding with future work in Section~5.

\section{Data and Current Methods}

Publicly available PITCHf/x data is available from several sources.\footnote{Our data was downloaded from \url{http://www.wantlinux.net/2009/10/pitch-fx-data-with-pitch-type/}. as well as data that we manually scrape from MLB.com with the aid of an edited Perl script and directions from the article \emph{How to Build a Pitch Database} [2]} Our data subset consists of pitches thrown by roughly 900 pitchers in the 2010 and 2011 seasons; we exclude data from before the 2010 season due to reported inconsistencies within the PITCHf/x system. One important point to make is that ground truth is not known, which is why we have measured stability of our method in Section 3.2, in terms of cluster memberships.  \\

The raw database contains trajectory information on each pitch, including acceleration and velocity, though not the spin of each pitch, which is useful in pitch classification because it can help distinguish between different pitch types. Some spin variables can be estimated using physics and a number of simplifying assumptions [3]; the name ``psuedo-spin'' is given to these quantities due to this. \\

There are several important variable definitions that we use throughout our paper and in our figures:

\begin{itemize}
\item The pitch's \textbf{start speed} measured in miles per hour at the release point.  This is measured using radar guns and is commonly acknowledged as the speed of the pitch.

\item The \textbf{back spin} of the baseball measured in radians per second.  A positive number represents back spin.  Most fastballs have back spin, while off-speed pitches have a tendency to have more ``top-spin'', or negative back spin.

\item The \textbf{side spin} of the baseball measured in radians per second.  A positive number represents left-to-right spin, or a left-handed pitcher's curveball or slider.  A negative number represents right-to-left spin, or the direction of a right-handed pitcher's curveball or slider.

\end{itemize}

Figure~1a plots these three variables for pitches thrown by Barry Zito; these pitches have been classified with the system developed via a neural network classification system for each pitcher developed by MLB Advanced Media.  Specifics pertaining to the method, model, and training data used are not publicly available [1]. \\



We have reason to believe Barry Zito threw 5 types of pitches in the 2010 and 2011 seasons due to the obvious 5 different clusters observed graphically, and according to \href{http://www.brooksbaseball.net}{Brooks Baseball}, whose PITCHf/x classifications are considered by some the most accurate available to the public, those pitches are a four-seam fastball, sinker, slider, curveball, and changeup; but they do not make their method or data available, only their results [3].  Moreover, we speculate that the pitches in the two-seam fastball cluster should instead be classified as sinkers since Brooks Baseball classifies these pitches as such.  The neural net classifies Brooks Baseball's sinker cluster as two-seam fastballs.  We investigate whether or not Barry Zito's two-seam fastball should be labeled as a sinker in Section~4 by implementation of a new classification algorithm. We ultimately agree with Brooks Baseball and label the cluster as a sinker.  Overall, our model tends to split up four-seam fastballs and two-seam/sinker clusters differently than the neural networks method in a way that more closely resembles Brooks Baseball classification and empirically makes more sense.  We explore this further in Section~4 when we discuss the Cliff Lee classification example. \\

In addition, the neural net classification appears to have obvious misclassifications.  Four-seam fastballs are classified as curveballs, and curveballs/changeups are classified as sliders, which are clearly labeled in the incorrect clusters in Figure 1a.  \\

\section{Model-Based Clustering with Gaussian Mixture Models}

Mixture models for clustering rely on a straightforward generative premise: there is a series of simple probabilistic models for how an event can be generated, and a weight on which model will be used to generate the observation. This description lines up directly with pitcher intent: while we as observers may not know what type of pitch is intended, the pitcher himself makes a choice of a specific pitch type (fastball, slider, curveball, etc) with a basic profile: a grip and arm motion that gives the ball a desired speed, spin and trajectory. \\

A multivariate Gaussian model for any particular pitch profile makes intuitive sense. Each coordinate has a mean value -- for example, a typical four-seam fastball might have an initial velocity of 95 miles an hour, a back spin of 100 radians per second and a side spin of 10 radians per second. The resulting pitch is then affected by many different sources of noise, both in the pitcher's delivery and in other external factors like the wind, and this noise can affect multiple pitch characteristics at once. The resulting pattern in three dimensions is an ellipsoid; in Figures 1b, 2, 3, and 5b, we visualize the MBC results for a variety of pitchers. One particular advantage to this approach is that we can detect when two clusters overlap, since geometry, and not just proximity, is an important factor.  \\

Given a set of Gaussian clusters and weights, this routine determines the probability that each observed pitch belongs to a given cluster as the relative probability density for the pitch if it were a member of each cluster, factoring in each cluster's relative weight -- most pitchers, for example, throw more fastballs than other off-speed pitches, and this is taken into account directly. In our classifications we declare a pitch to belong to the class with the highest probability under the model. \\

We use the \texttt{mclust} library in the \texttt{R} statistical programming software to perform our clustering operations, with some modifications that we describe to account for additional information. Given a pre-selected number of clusters, we use an Expectation-Maximization algorithm to calculate the maximum likelihood estimates (MLEs) for each cluster location, shape, and weight. Once we run this across a range of cluster counts, we use the Bayesian Information Criterion (BIC) model selection criteria to determine the optimal number of clusters in the model, which attempts to maximize the likelihood of the data while penalizing excessive numbers of parameters.\\

Empirically, MBC creates clusters of pitches that are more tightly confined than current pitch classifications, suggesting there are very few curveballs, changeups, or sliders misclassified. A major concern is the difference between the four-seam fastball and the sinker/two-seam fastball clusters, as seen in the Figures section.   We expect two-seam fastballs/sinkers to have a slightly slower start speed than four-seam fastballs because it is known four-seam fastballs to be a pitchers fastest pitch, but this is not the case with the PITCHf/x neural networks classification method.  However, MBC weights the velocity as an important factor between the two clusters and splits them accordingly. \\

Figure 1b shows the MBC as it performs for the pitches of Barry Zito; the red cluster is the faster pitch and light blue cluster is the slightly slower pitch with vertical spin closer to Zito's other off-speed pitches.  The MBC's potential four-seam fastball cluster (red) has the smallest number of pitches, but other sources suggest that Zito threw his four-seam fastball most often, leading us to suspect that while the labels (from Section 4) may need adjusting, the proper clusters are still being detected.  This also may indicate there is not much difference between Zito's four-seam and two-seam/sinker pitches from the batter's perspective.  To further support subtle advantages of MBC over the neural networks classification, we evaluate the overall stability of our clustering method in Section 3.2. \\

\subsection{Calculating the Number of Clusters}

In its original form, the model tends to choose more clusters than fewer for the pitchers we have tested. On inspecting the clusters produced, it is clear that the method is favoring relatively ``thin'' clusters, which have high internal correlations between variables, which is highly unrealistic for the physical examples we consider. Limiting the model to a smaller number of pitches is not generally feasible, and while manual inspection is possible, it would be far more preferable to automate the method to remove this issue. \\


We develop our own criterion for choosing the number of clusters, called the adjusted Bayesian Information Criterion, or BIC$_{\text{adj}}$. Since we observe that most pitch clusters are close to spherical, and our prior knowledge suggests that flat ellipsoidal clusters are unlikely, we are motivated to constrain the creation clusters which have high intra-cluster correlations between the three variables of interest.  Currently, each cluster $k$ has three parameter sets: $\mu_{k}$, the cluster mean; $\sigma_{k}$, the standard deviation of each dimension; and $\Sigma_{k}$, the intra-cluster correlation matrix. It is the terms of $\Sigma_k$ that need to be kept small in absolute value (1 or -1 indicates perfect correlation, 0 indicates no correlation). \\

To account for this, we develop an additional penalty term for the current BIC formula that adds a value proportional to each intra-cluster correlation term. Using BIC$_{\text{adj}}$, if the clusters the model finds with $k$=6 have high intra-cluster correlations, compared to $k$=5, than the correlation penalty term will be large, and BIC$_{\text{adj}}$ will be smaller for $k$=5.  We choose our $k$ based off of the minimum BIC or BIC$_{\text{adj}}$. \\

Figure 1b displays the clustering for Barry Zito chosen with adjusted BIC (BIC$_{\text{adj}}$), which chooses five clusters, which agrees with the pitch number selection of both the neural network classification and Brooks Baseball. For this application, BIC$_{\text{adj}}$ is a substantial improvement over BIC, and is the method we use going forward in choosing the number of pitch clusters. \\

In order to fully grasp all of the benefits the MBC method can offer, we investigate how it performs when clustering pitchers other than Barry Zito. In fact, we have investigated MBC's performance on the entire 2010 and 2011 season, and simply show a few illustrations from this large analysis performed. \\

Figure~2 shows how our method is ideal for detecting how a pitch can evolve over time.  The purple and black clusters represent Jon Lester's curveballs. The difference between the two clusters is the speed and horizontal break of the pitch.  In 2010, Jon Lester's curveball averaged 78 MPH while in 2011 it averaged 76 MPH with slightly more horizontal break -- a subtle but important difference.  Figure 3 visualizes how our method can also cluster pitches with high variances such as Tim Wakefield's knuckleball (orange); the pseudo-spins, corresponding to additional break, are considerably wider than for other pitches due to the unpredictable nature of stitch position. \\

\subsection{Stability Of Cluster Memberships}
We evaluate quality of the MBC method by assessing the {\em stability} of the method in terms of how sensitive our model is to smaller sample sizes.  We go about accomplishing this by taking the data for each pitcher and running the clustering on an 80\% subset of the data, the remaining 20\% subset, and then on the full data.  Next, we calculate the number of pitches in both the 80\% and 20\% subsets that do not change clusters compared to the full dataset.  In order to be confident in our results, we repeat this process 20 times for each pitcher and find the mean and standard error.  We find that after 20 samples our standard error is sufficiently small and we are confident with our stability sample mean estimates.  Figure 4 is the two distributions for the 80\% and 20\% subsets and the proportion of pitches that are in the same cluster as the full data for all pitchers.  Overall, for both the 80\% and 20\% subsets the majority of pitchers have 80\% or more of their pitches clustered in the same cluster.  We found that this stability holds on sample sizes as low as 100 pitches, or the average number of pitches a starting pitcher throws in one start.  It is important to note that we kept the number of clusters chosen (k) the same on both subsets as when we ran it on the full data. \\

\section{Classification}
In order to have a fully automated pitch clustering and classification system, we propose a simple classification algorithm, which improves pitch clustering by assigning sensible pitch labels to the respective clusters from our heuristic.  Although ground truth is unknown, we make comparisons with the current classification labels in the PITCHf/x database, Brooks Baseball classifications, and graphical visualization to determine how well our method works.  We found that our classification algorithm has many strengths including consistent performance and labeling multiple clusters as the same pitch type when appropriate, such as the Jon Lester curveball evolution over time scenario referenced in Section 3.2 and visualized in Figure~2.  Our clustering algorithm appears to have similar classification results as Brooks Baseball (which is the closest to the ``truth'' -- that is, a pitcher's actual choice of pitch to throw -- against which we can compare). \\

Our algorithm classifies and names a key subset of pitches: four-seam fastballs, two-seam fastballs, sinkers, change-ups, cut-fastballs, sliders, curveballs, and knuckleballs.  For each classification, the algorithm begins by assigning the cluster mean with the highest starting velocity as a four-seam fastball.  For each additional cluster, the algorithm goes through a series of constraints that are derived from each cluster's mean and variance to determine the label for each cluster.  For example, if a cluster mean has the same side spin direction as the four-seam fastball then it checks if the speed differs from the four-seam fastball by more than 6 MPH and if the side spin varies less than 60 rotations per second from the four-seam fastballs.  If it does, it assigns the cluster as a change-up.  If not, it determines if the side or back spin difference from the four-seam fastball is greater.  If the side spin difference is the greater of the two, it assigns the cluster as a two-seam fastball; if the back spin difference is larger, it assigns the cluster as a sinker.  The rest of the classification algorithm follows similar decision processes. \\

We evaluate how well our classification algorithm performs by taking a sample of various starting and relief pitchers with differing pitch repertoires.  We found after analyzing and comparing the results to Brooks Baseball and the neural network classification system, in 23 of the 25 scenarios our classification algorithm empirically appear correct and has similar classification results as Brooks Baseball.  The two pitchers that the algorithm does not classify correctly are Barry Zito and Derek Lowe.  It interchanges their sinker and four-seam fastball clusters because in the data sample, the measured speed of the sinkers are bigger than those for the four-seam fastball, a characteristic that is not commonly true for other pitchers. \\

In Figure 5, we visualize Cliff Lee's MLB-AM neural network classification compared to our method. Of particular note, our method separates and labels the two-seam and four-seam fastball by assigning the four-seam fastball to the fastest pitch with the least amount of back and top spin relative to Lee's other off-speed pitches.  Our method splits the two fastballs similar to Brooks Baseball, but Brooks Baseball instead labels the two-seam fastball as a sinker.  We also are able to observe that our method's slider (brown) cluster is clearly defined and classified with no other obvious misclassified pitches.  The neural network classification has obvious misclassification in the slider cluster with pitches labeled as changups, curveballs, and cut-fastballs.  Our method clearly improves both the clustering and classification of Cliff Lee compared to the neural network classification.  \\


\section{Summary and Future Work}

We have proposed a new clustering method and classification algorithm and tested both approaches using the PITCHf/x database for 2010-2011. Our analysis illustrates better clustering than the current neural network method based upon our plots and measuring the stability of our method. Furthermore, based upon our model based clustering method, we recommend a simple algorithm to classify each pitch based on the individual pitcher, which performs extremely well in most situations. Its strongest features are correcting any obvious misclassification in the neural network model, accounting for pitch evolution over time, and performing well for pitches with comparatively high internal variance.  Our method also performs well in a highly debated topic in MLB, namely distinguishing two-seam and four-seam fastballs. \\

These improvements suggest others can be made in assigning pitchers themselves to clusters, a problem highly motivated by the need to assess a pitcher's likely performance against unfamiliar players or teams. In these situations, most hitters likely have never faced the current pitcher, and thus, there is no or very little data to leading to inference about the hitter's history against the current pitcher. Placing pitchers into similar groups and looking for relationships with groups of batters would be invaluable to MLB. This information can lead to advanced and more detailed scouting reports of what type of pitches are a hitter's strength or weakness. Moreover, since we have prior information about many pitchers, we can utilize advanced Bayesian clustering methods and optimize the choice of $k$ for better and more stable results. 


\newpage
\noindent
{\bf References}\\
\noindent
\begin{enumerate}
\item[[1]]  Foster, Adam. {\it Scouting with PITCHf/x}, June 12, 2012. http://www.baseballprospectus.com/\\article.php?articleid=17327.
\item[[2]]  Fast, Mike. {\it How to Build a Pitch Database}, August 23, 2012. http://fastballs.wordpress.com/2007/08/23/\\how-to-build-a-pitch-database/
\item[[3]]  Nathan, Alan M. {\it The Physics Of Baseball} , accessed September 20, 2012. http://webusers.npl.illinois.edu/$\sim$a-nathan/pob/.

\end{enumerate}

\section{Appendix: Figures and Tables}

\begin{table}[h]
\begin{center}
{\scriptsize
\caption{Named pitch types with corresponding colors.}
\begin{tabular}{|c|c|c|c|c|c|}
\hline
{\bf Pitch Name}	&	Four Seam	&Two Seam &	 Sinker & Cut Fastball &	Changeup \\ \hline
{\bf Color}	&	Red & Grey	&Light Blue 	&Blue	&	Green	\\ \hline  \hline
{\bf Pitch Name} & Curveball	&	Knuckleball	&Slider		& Intentional Ball &	\\ \hline
{\bf Color}	&	Black and Purple	&	Orange	&	Brown	&	Yellow  &	\\ \hline
\end{tabular}
}
\end{center}
\end{table}

\begin{figure}[ht]
\begin{center}
\includegraphics[width=0.49\linewidth]{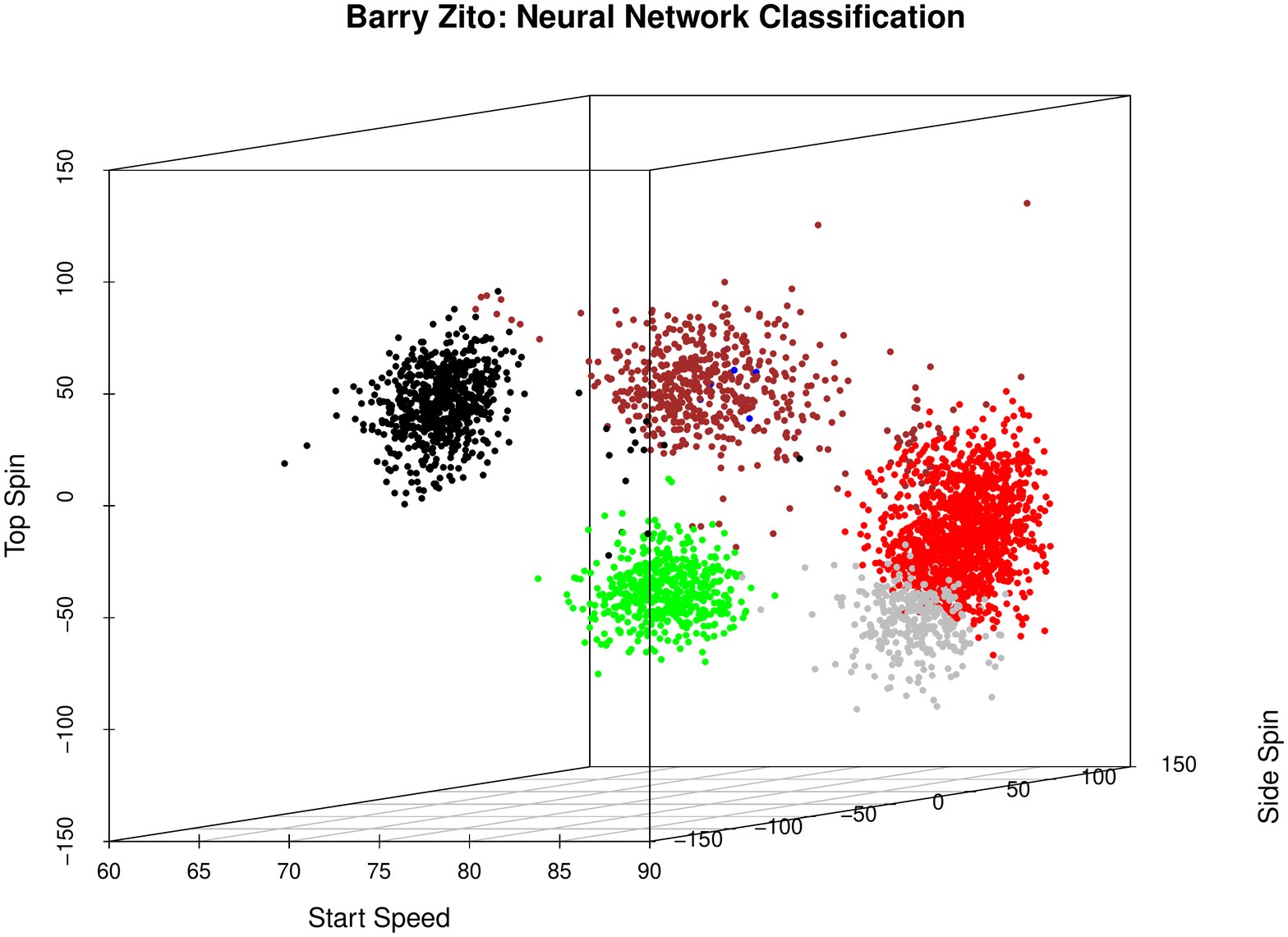}  \includegraphics[width=0.49\linewidth]{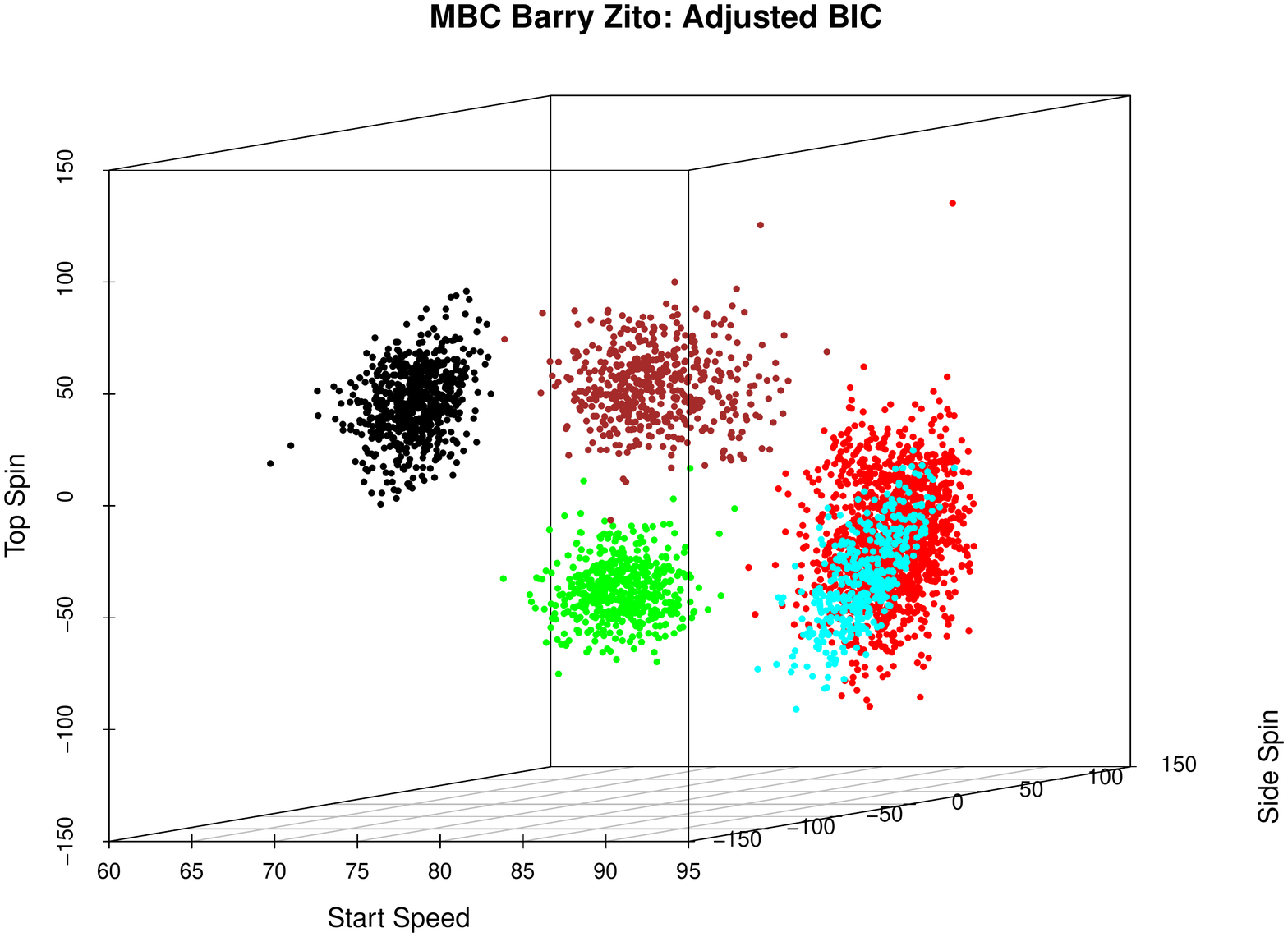}
\caption{\label{fig:ynplot-zito} The pitches thrown by Barry Zito. Figure 1a (left) displays the MLB Advanced Media classification developed via a neural network system.  There are obvious misclassifications: a small number of four-seam fastballs (which should be red) are classified as sliders (brown), as are some curveballs (black) and changeups (green).  Figure 1b (right) displays the MBC model and tightly clusters the pitches, as well as splits up the four-seam and sinker (light blue) clusters in an empirically sensible way.}
\end{center}
\end{figure}

\begin{figure}[ht]
\begin{center}
\includegraphics[width=0.5\linewidth]{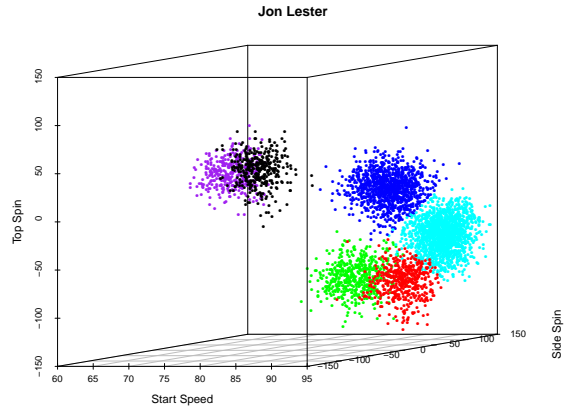}
\caption{\label{fig:ynplot-lester} The pitches thrown by Jon Lester classified using model-based clusters, which shows two distinct clusters for curveballs (purple and black) corresponding to a change over time. The difference between the two clusters is the speed and horizontal break of the pitch. This is one example of how the MBC model can detect subtle but important pitch evolution differences.}
\end{center}
\end{figure}

\begin{figure}[ht]
\begin{center}
\includegraphics[width=0.5\linewidth]{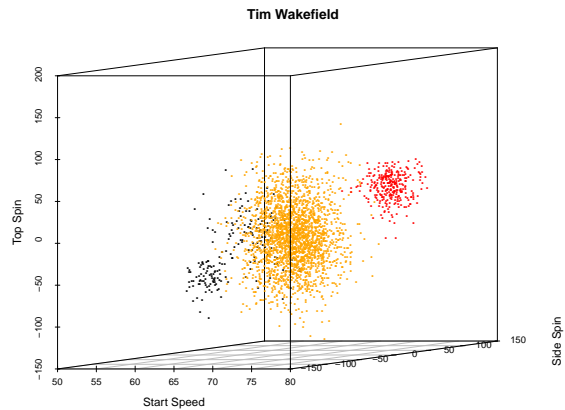}
\caption{\label{fig:ynplot-wake} The pitches thrown by Tim Wakefield classified using model-based clusters. The MBC model can also detect and classify pitches well with relatively higher variances like the pseudo-spins for Tim Wakefield's (spinless) knuckleball (orange), compared to the fastball (red) and curveball (black).}
\end{center}
\end{figure}



\begin{figure}[ht]
\begin{center}
\includegraphics[width=\linewidth]{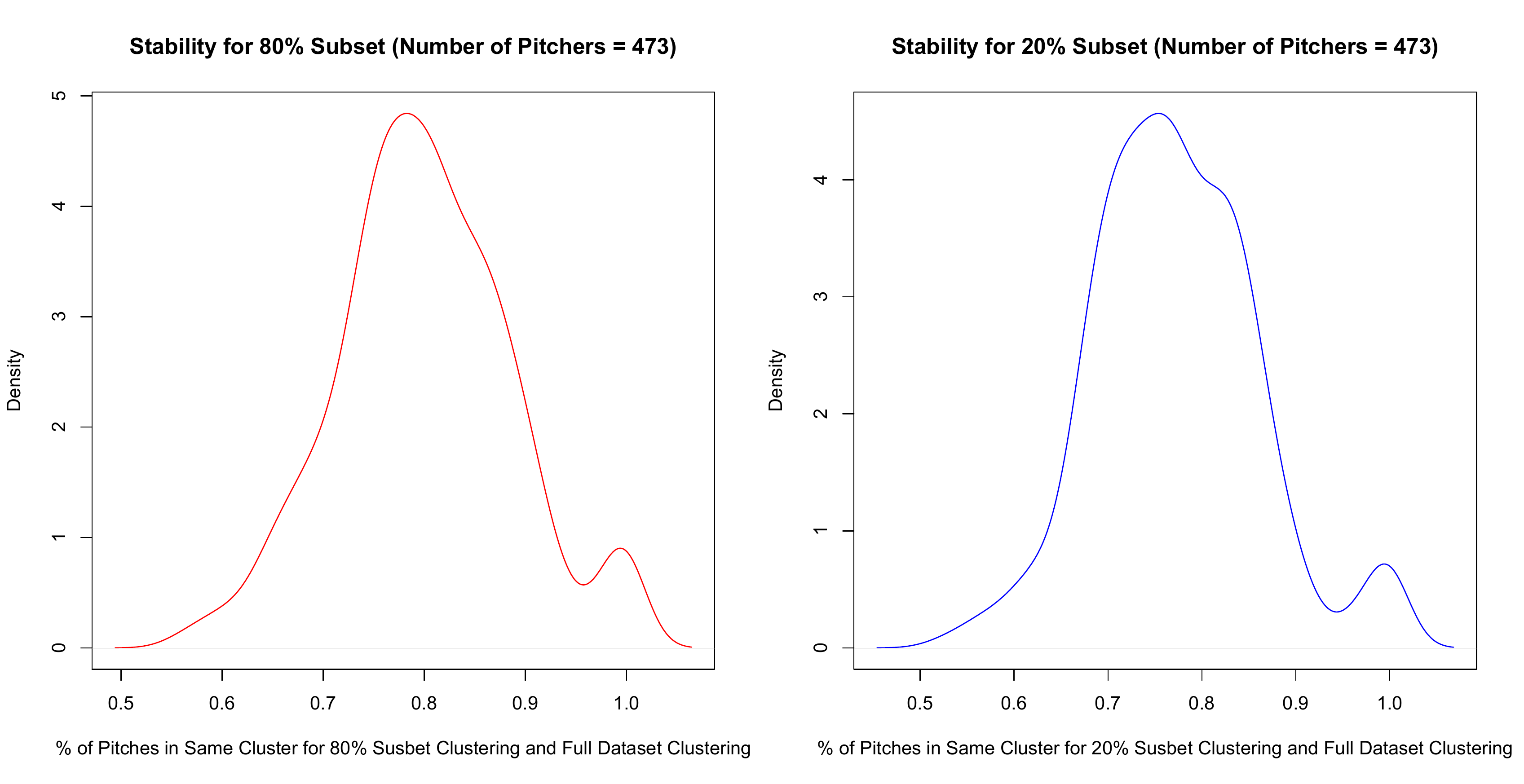}  
\caption{\label{fig:ynplot-stability} 80\% and 20\% Stability: Percent of pitches in the same cluster in subset and full dataset, across 20 replications of the procedure.  For both the 80\% and 20\% subsets the majority of pitchers have 80\% or more of their pitches clustered in the same cluster. We also found that this stability holds on sample sizes as low as 100 pitches, or the average number of pitches a starting pitchers throws in one start.}
\end{center}
\end{figure}

\begin{figure}
\begin{center}
\includegraphics[width=0.49\linewidth]{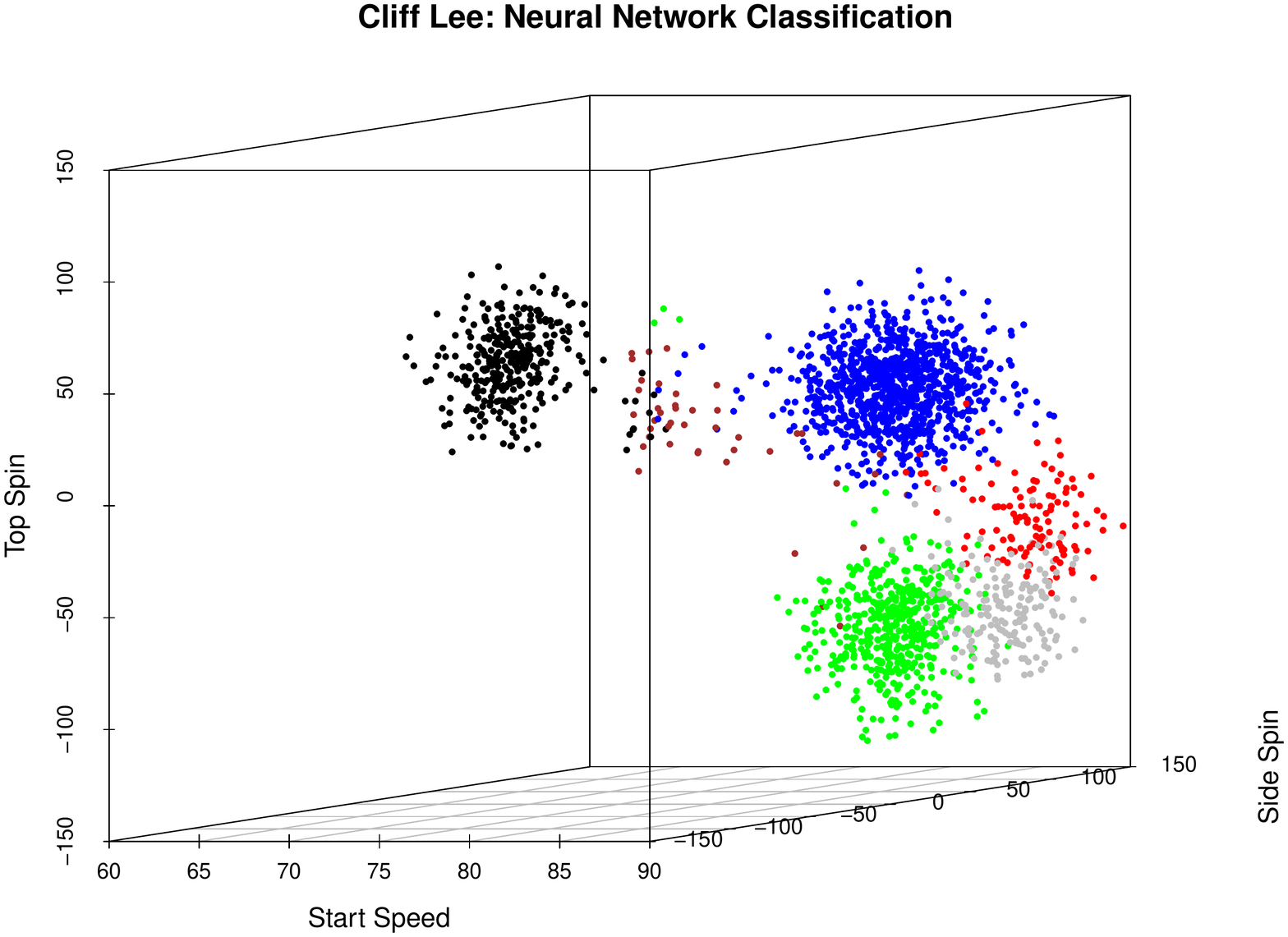}  \includegraphics[width=0.49\linewidth]{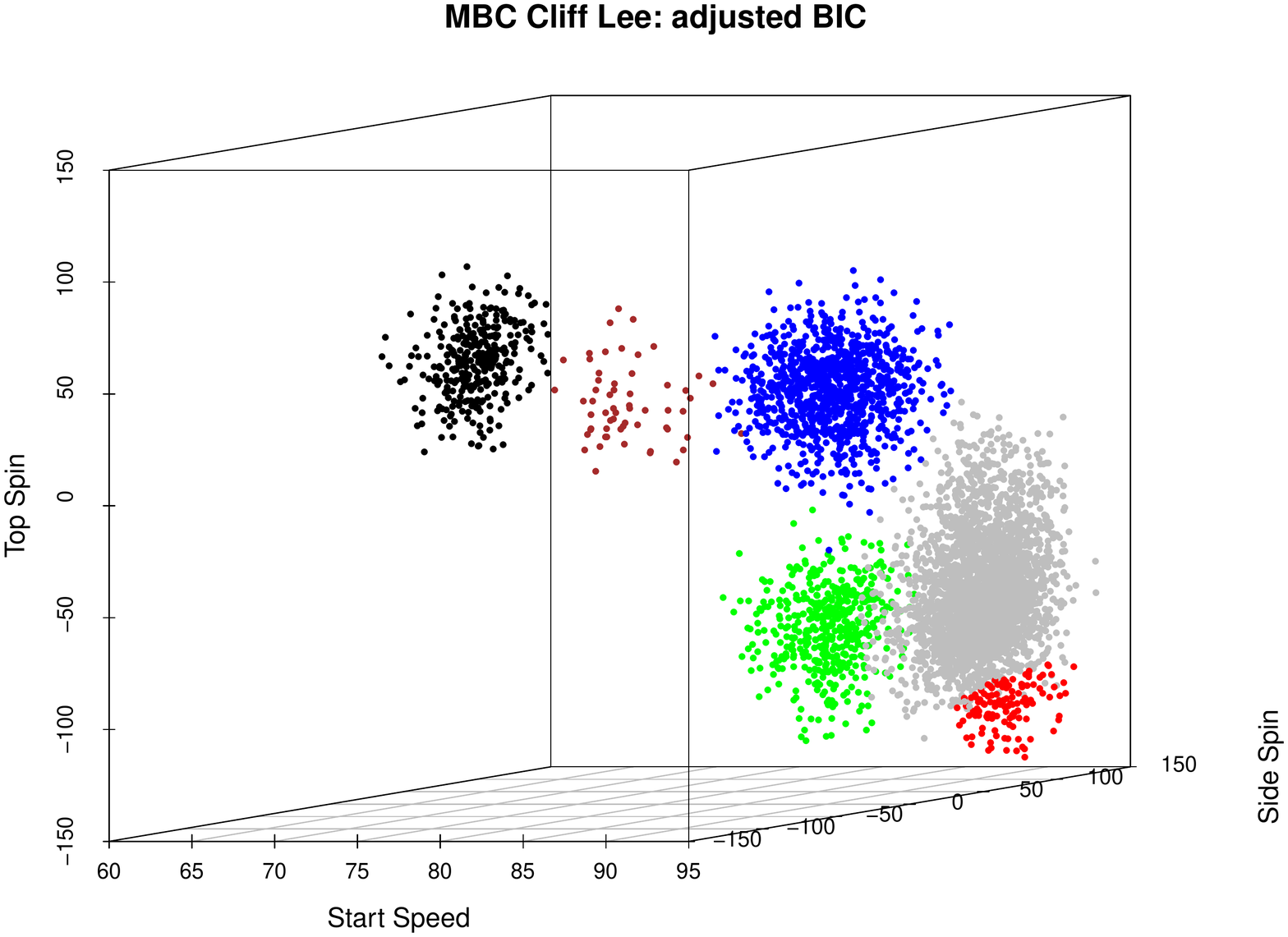}
\caption{\label{fig:ynplot-lee}Cliff Lee: Figure 5a (left) displays MLB's classification system developed via a neural network classification system.  As before, there are obvious misclassifications, but our method`s clusters are clearly defined and classified with no other obviously misclassified pitches. The MBC model in Figure 5b (right) separates and labels the two-seam (grey) and four-seam (red) fastballs by assigning the four-seam fastball to the fastest pitch with the least amount of back and top spin relative to Lee`s other off-speed pitches. This method agrees with the manually corrected data from Brooks Baseball.}
\end{center}
\end{figure}

\end{document}